\documentclass[aps,prb,twocolumn,superscriptaddress,floatfix]{revtex4-1}

\usepackage{bm}        
\usepackage{amsmath,amssymb,amsthm}   
\usepackage{mathtools}
\usepackage{braket}
\usepackage{tabularx}

\usepackage{hyperref}

\usepackage{grffile}
\usepackage{graphicx}
\usepackage{float}

\usepackage[version=4]{mhchem}
\newcommand*{\citen}[1]{%
  \begingroup
    \romannumeral-`\x 
    \setcitestyle{numbers}%
    \cite{#1}%
  \endgroup   
}

\begin{document}

\title{Terahertz Dielectric Analysis and Spin-Phonon Coupling in Multiferroic \ce{GeV4S8}}
\author{Matthew T. Warren}
\affiliation{Center for Emergent Materials, Department of Physics, The Ohio State University, Columbus, OH, USA}
\author{G. Pokharel}
\affiliation{Department of Physics and Astronomy, University of Tennessee, Knoxville, TN, USA}
\affiliation{Quantum Condensed Matter Division, Oak Ridge National Laboratory, Oak Ridge, TN, USA}
\author{A. D. Christianson}
\affiliation{Quantum Condensed Matter Division, Oak Ridge National Laboratory, Oak Ridge, TN, USA}
\affiliation{Department of Physics and Astronomy, University of Tennessee, Knoxville, TN, USA}
\author{D. Mandrus}
\affiliation{Department of Physics and Astronomy, University of Tennessee, Knoxville, TN, USA}
\affiliation{Department of Materials Science and Engineering, University of Tennessee, Knoxville, TN, USA}
\affiliation{Materials Science and Technology Division, Oak Ridge National Laboratory, Oak Ridge, TN, USA}
\author{R. Vald\'es Aguilar}
\email{valdesaguilar.1@osu.edu}
\affiliation{Center for Emergent Materials, Department of Physics, The Ohio State University, Columbus, OH, USA}
\date{\today}

\begin{abstract} 
We present an investigation of the multiferroic lacunar spinel compound \ce{GeV4S8} using time-domain terahertz spectroscopy. We find three absorptions which either appear or shift at the antiferromagnetic transition temperature, T$_N = 17$ K, as S=1 magnetic moments develop on vanadium tetrahedra. Two of these absorptions are coupled to the magnetic state and one only appears below the N\'{e}el temperature, and is interpreted as a magnon. We also observe isosbestic points in the dielectric constant in both the temperature and frequency domains. Further, we perform analysis on the isosbestic features to reveal an interesting collapse into a single curve as a function of both frequency and temperature, behavior which exists throughout the phase transitions. This analysis suggests the importance of spectral changes in the terahertz range which are linear in frequency and temperature.
\end{abstract}

\maketitle
\section{Introduction}

The lacunar spinel family of materials (\ce{AM4X8}) hosts many compelling phases of matter, including an insulating, N\'{e}el-type skyrmion,\cite{kezsmarki2015neel, ruff2015multiferroicity} superconductivity under pressure, \cite{abd2004transition} heavy-fermion behavior, \cite{rastogi1983itinerant,rastogi1987magnetic} and theoretically predicted two-dimensional topological insulation. \cite{kim2014spin} In these materials, spin-orbit coupling and electronic correlations can play a pivotal role. The lacunar spinel crystal structure differs from the typical spinel (\ce{AM2X4}) as every second A-site is removed, resulting in an \ce{NaCl} arrangement of tetrahedral ($\ce{AX4})^{n-}$ and cubane $(\ce{M4X4})^{n+}$ structures (see figure 1 in reference \citen{muller2006magnetic}). The tetrahedra can support magnetic moments and act as hopping centers. \cite{sahoo1993evidence,rastogi1996transport} Further, intracluster electronic arrangement, Coulombic interactions, and intercluster distances all support the insulating state, resulting in a novel type of Mott insulator. \cite{johrendt1998crystal, muller2006magnetic}

The skyrmion-host \ce{GaV4S8} is perhaps the most similar within this family to the material studied here, \ce{GeV4S8}. Both materials undergo Jahn-Teller driven structural transitions from the cubic phase in the 30-45 K range and magnetic transitions in the 10-20 K range. \cite{hlinka2016lattice,pocha2000electronic,ruff2015multiferroicity,yadav2008thermodynamic,sahoo1993evidence,muller2006magnetic,chudo2006magnetic,bichler2008structural,singh2014orbital,widmann2016structural}
While much of the magnetism in these compounds, including the novel insulating skyrmionic state in \ce{GaV4S8}, is driven by ferromagnetic (FM) interactions,  \ce{GeV4S8} is antiferromagnetic (AFM). \cite{yadav2008thermodynamic,chudo2006magnetic}

Prior work on \ce{GeV4S8} has revealed a multitude of couplings that drive the two phase transitions. A Jahn-Teller distortion (T$_{JT} = 31.5$ K) drives a symmetry change from a high-temperature, cubic F$\overline{4}$3m structure to orthorhombic, Imm2 ferroelectric (FE) state. \cite{bichler2008structural} We note that recent work has suggested that the high-temperature, paraelectric phase is actually the (perhaps dynamically appearing) tetragonal space group I$\overline{4}$m2. \cite{cannuccia2017combined} Orbital ordering also occurs at T$_{JT}$, where charge is reorganized within the vandium tetrahedral cluster.\cite{singh2014orbital} The lower AFM transition (T$_N = 17$ K) results from the coupling between the two unpaired electrons (S=1) that reside in the V4 clusters. \cite{johrendt1998crystal} Spin-phonon coupling has been previously reported in this material, as the tetrahedra are known to distort at T$_N$. \cite{bichler2008structural}

As this material undergoes FE and AFM transitions, it is classified as a multiferroic. Further, as ferroelectricity and antiferromagnetism occur at distinct temperatures, \ce{GeV4S8} is a type-I multiferroic. \citet{singh2014orbital} note that \ce{GeV4S8} displays an unusual combination of strong magnetoelectric coupling with distinct electric and magnetic ordering temperatures, making it a candidate for interesting device physics.\cite{singh2014orbital}

We use THz spectroscopy to study \ce{GeV4S8} since it has been a powerful tool in the investigation of the electrodynamic properties of novel magnetic materials in the last years. It has led to the discovery of new elementary excitations, \cite{Pimenov,Sushkov125,Mai2016terahertz} electromagnons in multiferroics, as well as to the furthering of our understanding of spin-phonon coupling in frustrated magnets.\cite{Sushkov124,Hemberger} In this work, we find absorptions coupled to the magnetic state in the THz frequency range. We also provide evidence for the observation of a magnetic absorption. Lastly, we investigate two parameter-independent points in our data, known as isosbestic points, with a technique from \citet{greger2013isosbestic} to reveal intrinsic parameter dependence and collapsed behavior, which in one case persist through the phase transitions.

\section{Experimental Methods}

Polycrystalline \ce{GeV4S8} was synthesized by solid state reaction. Stoichiometric amounts of germanium (99.999$\%$), vanadium (99.5$\%$) and sulfur (99.9995$\%$) were ground under an inert atmosphere and sealed in a silica tube. The tube was heated slowly to 750$^{\circ}$C and held at that temperature for 24 hours. The mixture was then ground again in an inert atmosphere and pressed into a pellet. This pellet was sealed in a silica tube and then heated to 800$^{\circ}$C for 20 hours. The phase purity of the powder was checked with powder X-ray diffraction and no signs of an impurity phase were found.

\begin{figure}[!t]
\includegraphics[width=1\columnwidth]{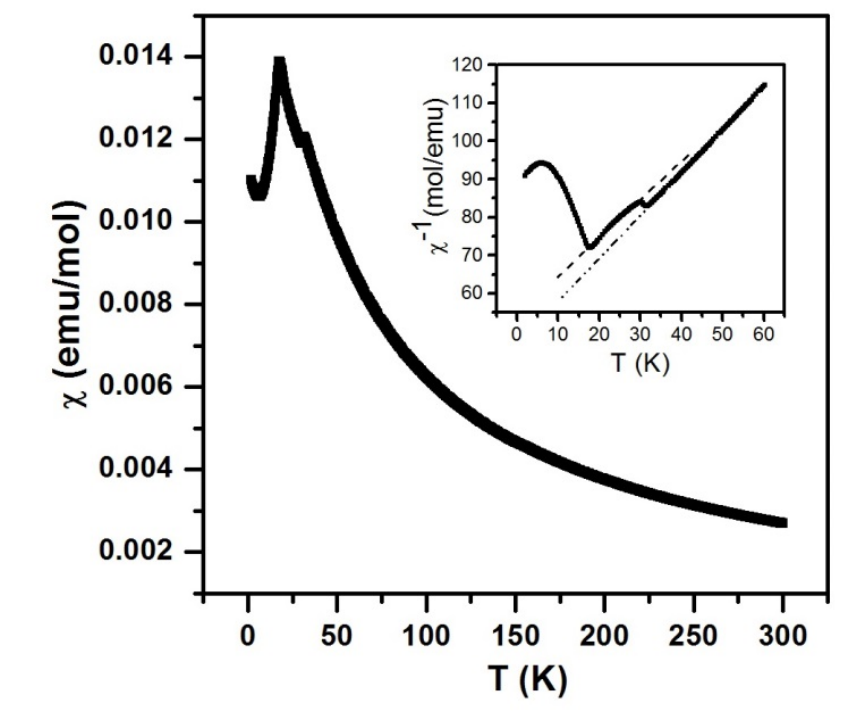}
\caption{Susceptibility measured upon warming under a 500 Oe applied field. The inset represents the inverse susceptibility at low temperatures. The dashed lines in the inset show Curie-Weiss fits in the region above and below the structural transition.}
\label{chi}
\end{figure}

Magnetic measurements were carried out in a SQUID magnetometer from Quantum Design from 2 K to 300 K. The magnetic susceptibility and low temperature inverse magnetic susceptibility are plotted in figure \ref{chi}. After zero-field cooling, the measurements were made upon warming from 2 K to 300 K under an applied magnetic field of 500 Oe. The susceptibility shows Curie-Weiss behavior above the 31.5 K structural transition. The anomalies seen at temperatures 31.5 K and 17 K represent the structural and magnetic transitions, respectively. The 17 K AFM transition appears as a cusp in the susceptibility curve. The observed N\'eel temperature is consistent with previous reports that have ranged from 13 to 18 K. \cite{muller2006magnetic, bichler2008structural,singh2014orbital} The step-like transition observed at 31.5 K has been shown to be the consequence of a structural transition from cubic (possibly tetragonal, as noted above) to orthorhombic. \cite{singh2014orbital} Above the structural transition at 31.5 K, the inverse susceptibilities are studied to discuss the Curie-Weiss behavior.

A Curie-Weiss fit from 40 K to 300 K yields an effective moment of 2.7 $\mu$B and a Curie-Weiss temperature of -44 K. A fit to the data between the transitions at 17 K and 31.5 K yields an effective moment of 2.86 $\mu$B and a Weiss temperature of -55 K. These observed Curie-Weiss parameters are comparable to previously reported values. \cite{widmann2016structural, bichler2008structural} The negative value of Curie-Weiss temperature is consistent with AFM exchange interactions. The fitted moments are consistent with spin-1 V$^{3+}$. 

Time-domain terahertz spectroscopy (TDTS)\cite{fattinger1989terahertz} was performed at temperatures down to 7.6 K, within frequency range $\sim$0.2 -– 1.5 THz. The high frequency limit is determined by the absorption of the cryostat windows (z-cut quartz).  In TDTS, a $\sim$20 fs pulse of 800 nm central frequency is split into two pulses via a beamsplitter and a delay stage is used to create a path length difference between the two pulses. One pulse travels to a biased photoconductive THz emitter and the other to a non-biased photoconductive THz detector. At both the emitter and detector, carriers of $<$1 ps lifetime are created. The generated carriers at the emitter accelerate due to the bias, creating a current which emits THz radiation into free space. After this THz pulse is focused through a sample via off-axis parabolic mirrors, it arrives at the photoconductive detector. The electric field of the transmitted THz pulse generates a current in the photoconductive detector, which is sent to a pre-amplifier and lock-in detector. This sectional measurement of the THz pulse via optical path length variation thus constitutes a full characterization of the electric field of the THz pulse in the time-domain. 

The complex transmission is experimentally determined by taking the ratio of the Fourier transforms of a pulse transmitted through the sample and a pulse transmitted through an empty aperture. Since we are measuring a pressed powder and thus all crystal orientations at once, we obtain pseudo-optical constants from the transmission coefficient. However, we omit the prefix \emph{pseudo} in the following. The powder measurement allows us to see all polarization/magnetization dependent absorptions without needing access to multiple single crystals cut along different planes.

In a slab geometry, the transmission is calculable from the Fresnel coefficients at normal incidence. The resulting equation is numerically solved for the complex index of refraction, $n(w) = n + i k$, and the index may then be converted to a frequency-dependent conductivity ($\sigma(\omega) = \sigma_1(\omega) + i \sigma_2(\omega)$) or dielectric constant ($\varepsilon(\omega) = \varepsilon_1(\omega) + i \varepsilon_2(\omega)$). As the conductivity and dielectric constant are related by $\varepsilon(\omega)= 1 + 4 \pi i \sigma(\omega)/\omega $, we present the dissipative optical constant $\sigma_1(\omega)$ and the dissipationless term $\varepsilon_1(\omega)$, resulting in a full description of the linear optical response of the material.

\section{Results and Discussion}

\subsection{Optical Constants}


\begin{figure*}[ht]
\includegraphics[width=1\textwidth]{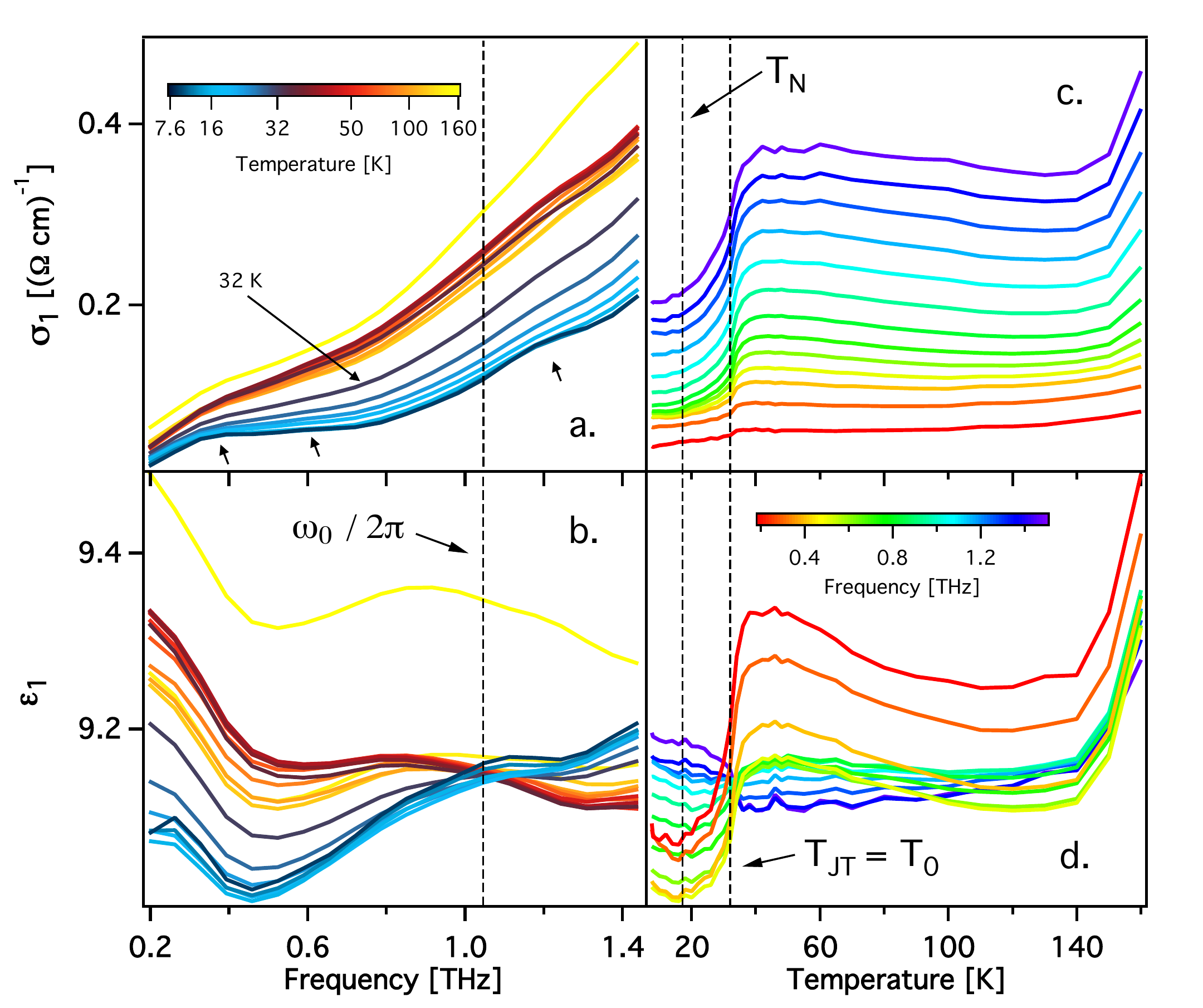}
\caption{The optical constants $\sigma_1$ and $\varepsilon_1$ are shown as a function of frequency (panels a. and b., respectively). Temperature cuts are taken of this data for panels c. and d. An isosbestic point is observed at 1.04 THz in b. and at $\sim$31.5 K in d. Vertical dashed lines mark the isosbestic frequency in a. and b. and mark T$_N$=17 K and T$_{JT}$=31.5 K in c. and d. Arrows in $\sigma_1$ mark the absorptions. Only a subset of measured data are shown, in particular, $\sigma_1(\omega)$ at 32 K, just above T$_{JT}$, is marked in a.}
\label{Eps1Sigma1}
\end{figure*}

In figure \ref{Eps1Sigma1}a., we notice several features in the real part of the conductivity, $\sigma_1$, versus frequency. A strong low-frequency absorption around 350 GHz, and weaker features at $\sim$600 GHz (which only appears below T$_N$) and $\sim$1.2 THz (see also figure \ref{SW}) are clearly observed. These features appear, shift, and/or sharpen below T$_N$, suggesting that they are either magnetic or couple to the magnetic state. The most prominent feature, at 350 GHz exists above and below both T$_{N}$ and T$_{JT}$, as does the 1.2 THz absorption. Importantly, as these two absorptions are present above T$_N$, they are not magnons. Also, all three absorptions sit on a background of increasing conductivity with frequency, which is interpreted as the contribution from the low-frequency tails of all THz-active phonons whose resonant frequency is above the available bandwidth. The effect of these phonons is observable as we are measuring a pressed powder and viewing all crystal orientations at once. Finally, we note an overall jump in $\sigma_1$ at T$_{JT}$ which will be discussed further below.
 
In figure \ref{Eps1Sigma1}b., we examine the real part of the dielectric constant, $\varepsilon_1$, versus frequency, and note the following. First, there is a low-frequency, low temperature feature which corresponds, via Kramers-Kronig relations, to the peak seen around 350 GHz in $\sigma_1$. Secondly, upon cooling, there is an abrupt decrease in the low-frequency ($\lesssim$ 1 THz) value of $\varepsilon_1$ below the Jahn-Teller structural transition, perhaps due to reduced fluctuations of polarization at the onset of FE order. Lastly, there is a frequency (marked $\omega_0/2\pi=$1.04 THz) where the value of $\varepsilon_1$ remains mostly unchanged throughout both phase transitions. We refer to this as a temperature-independent isosbestic point.

In figure \ref{Eps1Sigma1}c., which displays the constant frequency cuts of figure \ref{Eps1Sigma1}a., we present $\sigma_1$ vs. T, in which we similarly note the jump in conductivity as temperature is increased past the Jahn-Teller transition, indicating that THz conduction is more dissipative in the higher-temperature structure. A qualitatively similar jump is seen in \ce{GaV4S8}. \cite{wang2015polar} The prominent change in $\sigma_1$ observed at high frequencies in our data for \ce{GeV4S8} is believed to result from the phononic behavior above the available bandwidth, similar to the \ce{GaV4S8} phonon behavior observed in ref. \citen{hlinka2016lattice}.

In figure \ref{Eps1Sigma1}d., which contains the constant frequency cuts of figure \ref{Eps1Sigma1}b., a nearly frequency-independent isosbestic point is clearly seen in $\varepsilon_1$ at the Jahn-Teller transition. Also, we note a minimum at the first temperature we measured below T$_N$ (16 K), for frequencies below $\sim$500 GHz, which is the region where the low-frequency absorption is expected to have most contribution. This is comparable with the behavior of $\varepsilon_1$ through T$_N$ in \ce{GaV4S8} where there is a change in slope. \cite{wang2015polar} Above $\sim 140$ K, we see an increase in both $\varepsilon_1(\omega)$ and $\sigma_1(\omega)$ which is unexpected as there is no known phase transition in \ce{GeV4S8} at this temperature.

\subsection{Spin-Phonon Coupling}

\begin{figure*}[!ht]
\includegraphics[width=2\columnwidth]{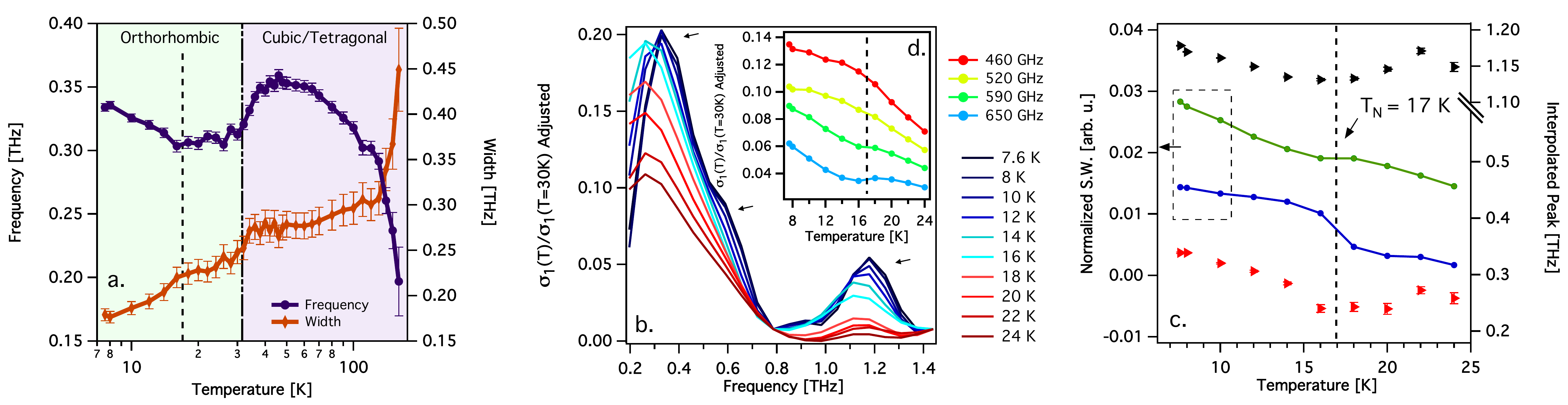}
\caption{The temperature dependence of the width and center frequency of the 350 GHz phonon is shown in panel a. The effects of spin-phonon coupling are clear in the changes of the behavior of the frequency of this mode at the magnetic ordering transition. Panel b. shows the ratio $\sigma_1(\text{T}<30K)/\sigma_1(\text{T}=30K)$, which emphasizes the temperature behavior of the three observed absorptions. We have adjusted this ratio as described in the text to isolate the response of the absorptions from the background. The absorptions are seen to occur around 350 GHz, 600 GHz, and 1.2 THz, as indicated by arrows. The bump at 900 GHz is an artifact of the analysis and not a feature. Panel c. shows the normalized spectral weight for the 520-720 GHz (\protect\includegraphics[scale=0.45]{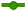}, multiplied by 2) and .85-1.44 THz (\protect\includegraphics[scale=0.45]{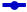}) regions and also the interpolated peaks from panel b. (\protect\includegraphics[scale=0.45]{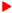}: 350 GHz absorption; \protect\includegraphics[scale=0.45]{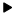}:  1.2 THz absorption; error bars are from Gaussian fits to $\sim\pm$ 100 GHz around each peak). The N\'{e}el temperature is marked by a vertical dashed line to emphasize the significance of the magnetic transition. Panel d. shows selected fixed frequency cuts of panel b. The emergence of the magnon is clearly seen below the N\'{e}el temperature marked as a vertical dash line.}
\label{SW}
\end{figure*}

As we observe all crystal orientations at once, modeling the absorptions at 350 GHz, 600 GHz and 1.2 THz with traditional forms can be suggestive, but not exact. Rigorous modeling can be especially challenging when separating the response of features in the bandwidth from the tails of unseen features outside of our bandwidth. In fact, without knowing the parameters of the relevant high-frequency phonons, it is essentially impossible to distinguish an inherently asymmetric phonon from a symmetric phonon whose apparent asymmetry is introduced by the increasing background. We, however, address this problem of parameter extraction in two ways: first, we make gaussian fits to the 350 GHz peak observed in $\varepsilon_2$ ($\propto \sigma_1/\omega$). We are here primarily concerned with two results: the gaussian center frequency and width. These results are shown in \ref{SW}a. and allow us to analyze frequency shift of the absorption, as well as sharpening. 

We find the expected anharmonic behavior of a phonon hardening upon cooling in the center frequency at high temperatures in figure \ref{SW}a. Around T $\sim$ 50 K~$\sim$ 1.5 T$_{JT}$, we begin to see softening of this excitation. This is possibly due to magnetic fluctuations coupling to the phonon, similarly to what has been observed in other systems \cite{Sushkov124,Hemberger,RolandoTb125,RolandoCdCr2O4,RolandoEuY}. Below T$_{JT}$ the frequency flattens, suggesting that the effect of these magnetic fluctuations has been reduced due to a transition to a new crystal structure. In this picture, finally, magnetic ordering ends the fluctuation-based softening and standard phonon hardening resumes below T$_N$. 

We make special note of the relation between our 350 GHz phonon and those observed in the infrared by \citet{cannuccia2017combined}. In particular, we draw comparison to the 325 cm${^{-1}}$ and 455 cm${^{-1}}$ ($\sim$9.7 and $\sim$13.6 THz, respectively) phonons, where these phonon frequencies are seen to increase when cooling to about 50 K, below which the phonons begin to soften. These modes are attributed to V4 intra-cluster bond length changes. Below T$_{JT}$, the frequencies remain temperature independent down to their lowest measured temperature of 22 K, as does our 350 GHz absorption. This is in complete agreement with our temperature-dependent modeling of the 350 GHz peak, and strongly suggests that the absorption we measure is related to the V4 cluster dynamics.

In the width of the gaussian fits we see evidence for near monotonic phonon sharpening. There are three clear regions of sharpening defined by the transition temperatures. This further emphasizes that both transitions govern the behavior of this mode. It has been suggested that the 350 GHz absorption is restricted to the V4 clusters, perhaps as a breathing-type mode, or an electronic/many-body excitation.\cite{reschke2017excitations} As the frequency of this absorption is seen to shift upon the phase transitions, where the lattice is known to distort, we find all of these explanations to convey the same message of a THz frequency absorption sensitive to spin and lattice dynamics. However, the strong similarity between the temperature dependence of the 350 GHz mode and the infrared phonons observed by \citet{cannuccia2017combined}, clearly indicates that this is a phonon and not an electronic excitation. In addition, these authors also suggest that the high temperature crystallographic structure is not cubic, but tetragonal as pointed out above. They find that distinguishing between these symmetries is extremely hard using scattering experiments. It is therefore possible that the 350 GHz phonon we observe is a signature of the structure being tetragonal and not cubic, as the lowest phonon in the cubic structure is not expected to have such a low frequency.

Our second approach to isolating the absorptions from the background is done by computing the ratio of $\sigma_1(\text{T}<30K)$ to $\sigma_1(\text{T}=30K)$ and subtracting away a linear term (defined by the line that joins frequencies 0.79 and 1.44 THz; subtraction of a quadratic term shows the same qualitative results). Here, we lastly add a small uniform offset to keep all resulting curves positive. The resulting data for temperatures at and below 24 K are plotted in figure \ref{SW}b. We then examine the quantity $\int \sigma_1(\text{T})/ \sigma_1(\text{T}=30 K)$d$\omega$ associated to these features in figure \ref{SW}c. to understand the temperature-dependent strength of the absorptions. Although this integral is not strictly the spectral weight of these modes, it is closely associated to it, and we use this terminology in what follows. We find the same qualitative conclusions when using reference temperatures above T$_{JT}$ or using a subtraction technique to find the difference in conductivity between temperatures above and below T$_{JT}$, which shows the robustness of the conclusions. 

In figure \ref{SW}b., we again find evidence that this 350 GHz absorption sharpens with decreasing temperature and its peak undergoes an abrupt shift in frequency below T$_N$. Around 600 GHz, we note a feature that appears only below T$_N$, and as can be seen from figure \ref{Eps1Sigma1}, this feature is the weakest of the three observed. For these reasons, we suggest that it is a magnon. Its normalized spectral weight, shown in figure \ref{SW}c., increases in the region 0.52 - 0.72 THz. While this mode sits on the shoulder of the 350 GHz phonon, it clearly protrudes from it with decreasing temperature and thus is regarded as an additional feature. In figure \ref{SW}d. we show constant frequency cuts of the adjusted conductivity ratios to display the distinct behavior of the magnon from the shoulder of the 350 GHz phonon; only within the frequency range of the magnon does the conductivity increases abruptly. Finally, we observe a frequency increase of the peak of the 1.2 THz absorption below T$_N$, suggesting that this absorption is also coupling to the magnetic state. This coupling is further evidenced by the increase of spectral weight in the .85 - 1.44 THz region. This absorption is not strong enough to appear isolated as the 350 GHz absorption is, meaning that the same high-temperature analysis will not work as ambiguities in background become significant. Thus, we cannot say if the same softening below 50 K occurs for the 1.2 THz absorption. 


\subsection{Isosbestic Points}

\begin{figure}[b]
\includegraphics[width=1\columnwidth]{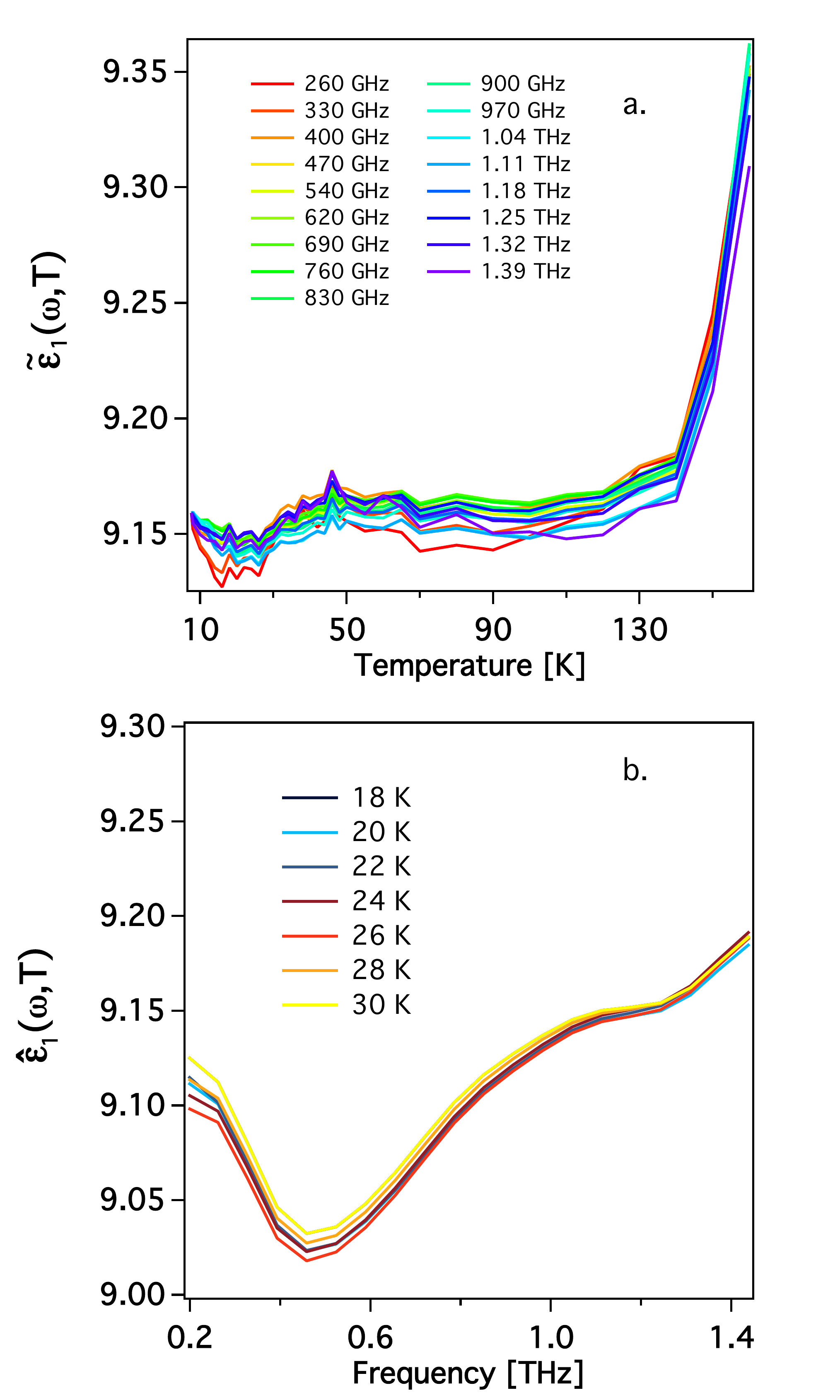}
\caption{Optical constant data collapse after performing isosbestic analysis \cite{greger2013isosbestic} on $\tilde\varepsilon_1$ in a., and $\hat\varepsilon_1$ in b. The collapsed behavior of $\tilde{\varepsilon}_1(\omega,\text{T})$ is nearly constant and persists across the phase transitions. The low temperature value has been subtracted from each $\tilde{\varepsilon}_1(\omega,\text{T})$ curve.}
\label{Iso}
\end{figure}

Isosbestic points occur when many curves of a parameter-dependent quantity display the same, or nearly the same, value at a fixed point of an independent variable. They are known to occur in correlated systems where data is taken over at least two different parameters, such as temperature, frequency, or doping. \cite{uchida1991optical, PhysRevB.81.235130, wang2016tuning, wang2014orbital} For example, as seen in this work, when there is a \textit{nearly} equal value of $\varepsilon_1$ for many frequencies at temperature T$_0$ (figure \ref{Eps1Sigma1}d., T$_0$=T$_{JT}$). These curves can be interpreted and analyzed in a manner similar to a Taylor-expansion around a central curve.\cite{greger2013isosbestic}  The nearly flat, $\omega_0/2\pi$=1.04 THz curve is found to be a valid central curve. When only the linear term is significant, this expansion takes the form (we refer to the measured $\varepsilon_1$ as $\varepsilon_1^{data}$ here to emphasize the role of the experimentally taken data which is displayed in figure \ref{Eps1Sigma1}):

\begin{equation}
\varepsilon_1^{data}(\omega,\text{T}) = \varepsilon_1^{data}(\omega_0,\text{T}) + (\omega-\omega_0) F_1(\text{T}) + \mathcal{O}[\omega^2]
\end{equation}

\noindent where

\begin{equation}
F_1(\text{T}) = \frac{\varepsilon_1^{data}(\omega_2,\text{T}) -– \varepsilon_1^{data}(\omega_1,\text{T})}{\omega_2 - \omega_1}
\end{equation}

\noindent and $\varepsilon_1^{data}(\omega_0,\text{T})$ refers to the curve for fixed frequency $\omega_0$ that is a function of temperature, as in figure \ref{Eps1Sigma1}d. We note $\omega_1$ and $\omega_2$ can be far removed from the frequency of the central curve, $\omega_0$. The exponent of the terms necessary in the expansion reveals the principal parameter dependence of the system, as in \ce{LaMnO3} where isosbestic analysis reveals a T$^2$ dependence of electronic excitations in the optical response without a linear-in-T term. \cite{PhysRevB.81.235130,greger2013isosbestic} Therefore this analysis can reveal the fundamental dependence of a physical quantity on measurement variables. For example, the analysis of our data shows that the behavior in $\varepsilon_1$ is linear in frequency and temperature.

We can obtain a collapse of the curves (again keeping only the linear power) by examining the function:

\begin{equation}
\begin{split}
\tilde{\varepsilon}_1(\omega,\text{T}) &\equiv \varepsilon_1^{data}(\omega,\text{T}) - (\omega-\omega_0) F_1(\text{T}) \\
&=  \varepsilon_1^{data}(\omega_0,\text{T}) + \mathcal{O}[\omega^2].
\label{isotilde}
\end{split}
\end{equation}

\noindent Assuming that second-order deviations from central behavior are small, the curve $\tilde{\varepsilon}_1(\omega,\text{T})$ for frequency $\omega$ will resemble the curve of the central frequency $\omega_0$, $\varepsilon^{data}_1(\omega_0,\text{T})$. Our convention will be that $\tilde{\varepsilon}_1(\omega,\text{T})$ (as in equation \ref{isotilde}) refers to the underlying behavior of the frequency-independent isosbestic point (as seen in figure \ref{Eps1Sigma1}d.), and we define $\hat{\varepsilon}_1(\omega,\text{T})$ as the underlying behavior of the temperature-independent isosbestic point (as seen in figure \ref{Eps1Sigma1}b.), with the roles of $\omega$ and T exchanged in the analysis:

\begin{equation}
\begin{split}
\hat{\varepsilon}_1(\omega,\text{T}) &\equiv \varepsilon_1^{data}(\omega,\text{T}) - (\text{T}-\text{T}_0) G_1(\omega) \\
&=  \varepsilon_1^{data}(\omega,\text{T}_0) + \mathcal{O}[\text{T}^2],
\label{eq:isohat}
\end{split}
\end{equation}

with 
\begin{equation}
G_1(\omega) = \frac{\varepsilon_1^{data}(\omega,\text{T}_2) -– \varepsilon_1^{data}(\omega,\text{T}_1)}{\text{T}_2 - \text{T}_1}.
\end{equation}

We first calculate $\tilde{\varepsilon}_1(\text{T})$ for all frequencies. We choose 1.04 THz as the central frequency becuse it is nearly temperature-independent, e.g. figures \ref{Eps1Sigma1}b. (crossing point) and 2d. (nearly flat line). The function $F_1(\text{T})$ is calculated using $\omega_1/2\pi =$ 0.473 THz and $\omega_2/2\pi =$ 1.25 THz. We emphasize that this choice is not unique in allowing the collapse of the data (in fact, surprisingly, every available frequency in the available bandwidth can serve as a central curve). We finally subtract away the difference of the low-temperature (7.6 K) value of $\varepsilon_1$ between each frequency and the isosbestic frequency, 1.04 THz. This allows us to reduce the vertical spread of the curves by a factor of 5 without changing their shape. The results are presented in figure \ref{Iso}a., for first-order subtraction, which provided the best results.

We have thus identified an intrinsic behavior of $\varepsilon_1$ by removing first-order deviations in frequency around the $\omega_0 / 2 \pi$ = 1.04 THz curve. This collapsed behavior is shown to be nearly frequency and temperature-independent, matching the behavior of $\varepsilon_1^{data}(2\pi*1.04$ THz, \text{T}) for all curves, as is clear from figure \ref{Iso}a. We thus show that there is an intrinsic behavior that exists at temperatures above and below T$_{N}$ and T$_{JT}$, which persists even up to high temperature ($\sim$160 K). This shows that the principal variations of the spectra in frequency are linear around the isosbestic frequency. Since this behavior is observed in $\varepsilon_1$, it is not related to conservation of spectral weight (as it would be if an isosbestic point was observed in $\sigma_1$). Although this collapse is similar to the scaling associated to a phase transition, it is unclear that this is the physical origin of this behavior; this collapse occurs without regard to the two known phase transitions in this material. At this moment it is unclear what the physical origin of this behavior is.

We find similar behavior in $\hat{\varepsilon}_1(\omega)$ only in temperatures between T$_{N}$ and T$_{JT}$. Again, removal of the linear term provided the best results as shown in figure \ref{Iso}b. The collapse only occurring between T$_N$ and T$_{JT}$ may imply that no new spectral features appear in this temperature range. This supports our view that the 600 GHz absorption only appears below T$_N$. Therefore, we have shown that $\varepsilon_1$ in the THz range in \ce{GeV4S8} has both a temperature-independent frequency, and a frequency-independent temperature, yielding a central isosbestic point at 1.04 THz and 31.5 K. While 31.5 K corresponds to the known phase transition temperature T$_{JT}$, the significance of the frequency 1.04 THz is not yet understood. 

We note that other materials also show isosbestic features in the THz range. \ce{GaV4S8}, which undergoes similar phase transitions to \ce{GeV4S8}, shows an isosbestic point just below T$_{JT}$. However, \ce{GaV4S8} does not show a temperature-independent frequency, as our data on \ce{GeV4S8} does (figures \ref{Eps1Sigma1}b. \& d.). Also displaying a different isosbestic behavior is the iron selenide superconductor \ce{Rb_{1-x}Fe_{2-y}Se_{2-z}S_{z}}, which shows isosbestic points in both $\sigma_1(\omega)$ and $\varepsilon_1(\omega)$ versus temperature but none versus frequency.

\section{Conclusions}

THz spectroscopy has proved sensitive to the antiferromagnetic and ferroelectric phase transitions in \ce{GeV4S8}. We have observed three absorptions in our frequency range: two clearly change below the antiferromagnetic transition, providing evidence for spin-phonon coupling, while the third we identify as a magnon as it appears only below the antiferromagnetic transition. We have also observed an isosbestic point in frequency and one in temperature, which occurs at the ferroelectric transition. By removing first-order terms in expansion around these isosbestic points, we are able to uncover underlying behavior in both observed isosbestic points, characterized by the collapse of $\varepsilon_1$ to a central curve. Further studies on single crystals are needed to resolve a number of outstanding questions raised by our studies.

Work at OSU was supported by the NSF MRSEC Center for Emergent Materials under Grant DMR-1420451. GP and DM were supported by the Gordon and Betty Moore Foundation's EPiQS Initiative through Grant GBMF4416. Work at the Spallation Neutron Source was supported by the Scientific User Facilities Division, Office of Basic Energy Sciences, U.S. Department of Energy. Evan Jasper, Shirley Li and Thuc T. Mai are thanked for their assistance in this work.

\bibliography{GeV4S8}


\end{document}